 \documentclass[showpacs]{revtex4}
\usepackage{fullpage}
\usepackage[dvips]{graphicx}
\input{epsf}
\usepackage{amsmath}
\usepackage{amssymb}
\usepackage[mathscr]{eucal}

\newcommand{\ket}[1]{| #1 \rangle}

\newcommand{\braket}[2]{\langle #1 | #2 \rangle}

\newtheorem{proposition}{Proposition}

\begin{document}

\title{On the generation of sequential unitary gates from continuous time Schr{\"{o}}dinger equations driven by external fields}
\author{Claudio Altafini}\thanks{This work was supported by a grant from the Foundation Blanceflor Boncompagni-Ludovisi.}
\affiliation{SISSA-ISAS  \\
International School for Advanced Studies \\
via Beirut 2-4, 34014 Trieste, Italy }
\email{altafini@sissa.it}


\begin{abstract}
In all the various proposals for quantum computers, a common feature is that the quantum circuits are expected to be made of cascades of unitary transformations acting on the quantum states.
A framework is proposed to express these elementary quantum gates directly in terms of the control inputs entering into the continuous time forced Schr{\"{o}}dinger equation.
\end{abstract}

\maketitle 



\section{Introduction}

In quantum information, the ``computing'' with quantum states is accomplished by applying sequences of discrete unitary gates, i.e. predetermined elements of the transformation group that acts on the state of the quantum system and determines its dynamical properties.
Such set of unitary transformations is normally specified in terms of the corresponding infinitesimal generators and, in order to guarantee arbitrary manipulation of the quantum states, it should be enough rich to generate the whole transformation group.

In quantum computing, this property is called universality of the gates \cite{Deutsch1} and in \cite{Lloyd1} it is shown that ``almost all gates are universal'', see also \cite{Ramakrishna4,Weaver1} for a control theoretic reformulation of the same idea.
The problem is treated also in \cite{DAlessandro4} by recollecting results on the uniform generation of the rotation groups.
The main result is the following: arbitrary state transfer is achievable by means of a number of unitary gates equal to the dimension of the transformation group (henceforth $ SU(N) $ of dimension $ N^2-1 $).
Such unitary gates do not have to correspond to linearly independent infinitesimal generators: in fact, owing to the semisimple nature of $ SU(N) $, almost all pairs of infinitesimal generators in the Lie algebra $\mathfrak{su}(N) $ allows to achieve arbitrary state transfer by cascading $ N^2-1$ exponentials of the two generators in alternate sequence.
For the single qbit, this is the idea behind for example the ZYZ parameterization of rotations of the corresponding group of transformations $SU(2)$.
For $N$qbit systems, parameterizations aiming at the same scope are proposed for instance in \cite{Murnaghan1,Reck1}.
From a technological viewpoint, this is a great simplification as it means that in practice arbitrary manipulation can be done by sequencing control pulses applied along (almost) any pair of laboratory fixed axes.
2D NMR spectroscopy is one field where such techniques of sequencing pulses from two different directions has reached the level of a very sophisticated science \cite{Ernst1,Mehring1}.
The cascade of unitary transformations obtained in this way is a product of exponentials, each exponential containing a parameter expressing the time duration of the pulse (scaled by a constant depending on the ``strength'' of the applied field).
By varying these $N^2-1 $ parameters, every quantum state is reached.

In general, the philosophy of control synthesis for states that evolve in continuous time governed by differential equations (like the Schr{\"{o}}dinger equation) is to look for input-parameterized solutions of the dynamics and then to select the control inputs in order to accomplish the given tasks (for example state transfer).
As mentioned above, in the sequential model for quantum computing it is normally expected that the elementary gates that constitute the ``quantum circuits'' will be made of discrete sets of unitary operators corresponding to fixed values of the $ N^2 -1 $ parameters mentioned above, applied in a fixed order.
In other words, what in a different context would be simply a parameterization of a group manifold, like a set of Euler angles, in the context of quantum computing becomes the ``hardware'' basis for the construction of quantum circuits.
This ``mimicking'' the behavior of classical circuits is motivated by the need to simplify as much as possible the influence of the dynamics in the logic of the circuits and of the ``hardware'' requirements of the circuits, but it is complicated by the continuum of values in which the quantum state can exist, as opposed to its classical counterpart (i.e. the bit).

The main scope of this paper is to provide a framework for reconciliating the continuous time forced Schr{\"{o}}dinger equation with the ``discrete dynamics'' of quantum computers.
The formalism used is that of the Wei-Norman formula, which relates the Magnus expansion (suitable to represent the continuous time evolution of the Schr{\"{o}}dinger equation) with the product of exponentials expansion corresponding to complete sets of elementary gates.
Essentially, both expansions define local coordinate charts on the group of unitary transformations by means of a basis of the corresponding Lie algebra and via the exponential map.
That is why these are sometimes referred to as {\em exponential coordinates} (or, in some cases, canonical coordinates).
For compact groups, the two corresponding parameter spaces are different: one is a solid sphere and the other a cube, and the two sets of coordinates are related by nonlinear differential equations.
An algorithm for the explicit computation of such differential equations (i.e. of the Wei-Norman formula) in terms of the structure constant of the Lie algebra was recently proposed in \cite{Cla-param-diff1}.
As for all series expansions on semisimple Lie groups, existence is not a global property, see \cite{Wei2}.
In fact also the Wei-Norman formula has a singular locus which is an algebraic set of the parameter space.

Another control-theoretically sound way to design inputs that resemble elementary quantum gates is to use piecewise constant controls \cite{Ramakrishna2,Schirmer4}. 
The methods proposed here could be seen as the differential version of the control design obtained via piecewise constant controls.
In fact, also the flow induced by piecewise constant controls looks like a product of exponentials which resembles a cascade of elementary gates.

For sake of comprehension, a parallel with what happens in the control of robotic chains \cite{Craig1} is useful to understand this relation: while at each time instant one can express the position of the end-effector of a robot in terms of joint angles and topology of the robot via a {\em static} map, called forward kinematics, (in quantum control: a map constructed from the Lie group decompositions of the unitary propagator, i.e. from products of exponentials), in order to describe dynamic changes it is convenient to pass to the differential forward kinematics i.e. to the {\em Jacobian} map obtained by differentiating the static map mentioned above. In Robotics, the use of differential maps is crucial whenever a closed form expression of the inverse of the forward kinematics (giving the values of the joint angles in terms of the end effector position) is not available. The Wei-Norman formula is exactly the Jacobian of the change of coordinates between single exponential and product of exponentials. Furthermore, as we will see, it automatically allows to bypass what is the main problem of the piecewise constant control methods, that is to say finding an explicit value of the $N^2 -1 $ parameters that constitute a decomposition of an element in $SU(N)$ for $N> 2$ (in the robotic case this would correspond to the inversion of the forward kinematic map). Lastly, while the Lie group decompositions used in piecewise constant control methods \cite{Ramakrishna3,Schirmer5} are not directly related to the positioning of the laboratory equipement that produces the control fields on the quantum system, this factor is implicitely considered in the product of exponentials we will use here.

\section{Exponential coordinates on $SU(N)$}
\label{sec:exp-coo-SUN}
Consider a closed finite level quantum system described by a state $ | \psi
\rangle$ evolving according to the time dependent Schr\"{o}dinger equation
\begin{equation}
\begin{split}
i \hbar  {\ket{\dot \psi(t) }} & = H(t) \ket{\psi} = \left( H_0 +  H_i(t)  \right) \ket{\psi(t) }  \\
\ket{\psi(0) } & = \ket{\psi_0 } 
\end{split} 
\label{eq:schrod1}
\end{equation}
where the state of the $N$-level quantum system $ \ket{\psi}$ lives on the sphere in $N$-dimensional complex Hilbert space $ \mathbb{S}^{N-1} = \left\{ \ket{\psi} \in \mathbb{C}^N \, \text{ s.t. } \, \braket{\psi}{\psi} = 1 \right\} $, and the traceless Hermitian matrices $ H_0 $ and $ H_1(t) $ are
respectively the constant internal Hamiltonian and the external time-varying
Hamiltonian, this last representing the interaction of the system with the control fields.
The solution of \eqref{eq:schrod1} is normally written in terms of the unitary propagator $ U(t) \in SU(N)$:
\[
\ket{\psi} = U(t) \ket{\psi_0} 
\]
with $U(t) $ satisfying an equation similar to \eqref{eq:schrod1} but lifted from the sphere $ \mathbb{S}^{N-1} $ to the special unitary group $SU(N) $:
\begin{equation}
\begin{split}
i \hbar \dot{U}(t)  & = H(t) U(t) = \left( H_0 +  H_i(t)  \right) U(t)  \\
U(0)  & = I
\end{split} 
\label{eq:schrod2}
\end{equation}
From now on, we shall use atomic units $\hbar=1 $.
Assume $A_1 , \ldots , A_n $, $ n=N^2 -1 $, are skew-hermitian matrices forming a basis of the Lie algebra $\mathfrak{su}(N) $ and that the Hamiltonian $H(t) $ can be written in this basis as 
\[
- i H(t) = -i \left( H_0 + H_i \right) = \sum_{j=1}^n a_j A_j + \sum_{j=1}^n u_j A_j 
\]
where $ a_j $ and $ u_j = u_j (t) $ are respectively the (constant) components of the free Hamiltonian along the basis directions and the (time-varying) control parameters of the interaction part, some of which might be zero if the control along the corresponding direction is missing.

If $T$ is the Dyson time ordering operator, 
the solution of \eqref{eq:schrod2} is formally written in terms of the exponential map as
\begin{equation}
U(t) = T { \rm exp}\left(- i  \int_{0}^t H(\tau ) d \tau \right) = T  { \rm exp}\left(- i  \int_{0}^t\sum_{j=1}^n (a_j  + u_j(t) ) A_j \right) 
\label{eq:Mag-exp1}
\end{equation}
normally referred to as Magnus expansion \cite{Magnus1}.
Alternatively, instead of \eqref{eq:Mag-exp1}, the solution of \eqref{eq:schrod2} locally admits an expression in terms of product of exponentials, first due to Wei-Norman 
\begin{equation}
U(t) = {\rm exp}(\gamma_1 A_1 ) \ldots {\rm exp}( \gamma_n A_n )
\label{eq:WeiN-exp1}
\end{equation}
with generally time dependent parameters $ \gamma_i = \gamma_i (t) $.
The relation between the two expansions \eqref{eq:Mag-exp1} and \eqref{eq:WeiN-exp1} is given by the so-called {\em Wei-Norman formula}, which expresses the functions $ \gamma_i (t) $ in terms of the $ a_i + u_i(t)$ via a system of differential equations:
\begin{equation}
\Xi (\gamma_1 , \ldots , \gamma_n ) \begin{bmatrix} \dot \gamma_1 \\ \vdots \\ \dot \gamma_n \end{bmatrix} =  \begin{bmatrix} a_1 + u_1 \\ \vdots \\ a_n + u_n \end{bmatrix} \qquad \gamma_i(0) = 0 
\label{eq:matrix-xi}
\end{equation}
with the $n\times n $ matrix $ \Xi $ analytic in the variables $ \gamma_i $.
Since the adjoint maps can all be written as (summation convention over repeated greek indexes): 
\begin{equation}
e^{\gamma_j {\rm ad}_{A_j} } A_i =
e^{\gamma_j A_j } A_i e^{ - \gamma_j A_j} = \sum^{\infty}_{l=0} \frac{
  (\gamma_j )^l }{l!}  {\rm ad}_{A_j}^l A_i = \sum^{\infty}_{l=0} \frac{
  (\gamma_j) ^l }{l!} c_{i \, j } ^{\mu_1} c_{i \, \mu_1} ^{\mu_2} \ldots c_{i \, \mu_{l-1} } ^{\mu_l }  A_{\mu_l} 
\label{eq:adj2}
\end{equation}
the matrix $ \Xi $ of elements $ (\Xi)_{ki}= \xi^k_i$ is defined in terms of the $ \gamma_i $ and of the structure constants as:
\begin{equation}
\prod_{j=1}^m e^{ \gamma_j {\rm ad} _{A_j } } A_i = \xi_i^\mu A_\mu  \qquad m=1, \ldots , n 
\label{eq:wei-tmp1}
\end{equation}
When the cardinality of the basis ordering is followed in \eqref{eq:WeiN-exp1}, the matrix $ \Xi $ assumes also the meaning of map between canonical coordinates of the first kind and canonical coordinates of the second kind, see \cite{Varadarajan1}.
In this case, since $ \gamma_i(0) =0 $, $ \Xi(0) = I $ and thus $ \Xi$ is locally invertible.
However, any fundamental parameterization of $SU(N) $ could be used: various possible sets of Euler-like angles for $SU(N) $ are discussed in \cite{Murnaghan1,Reck1,Nemoto1}.
For example, we will concentrate on the ZYZ parameterization of $SU(2)$ in the example of Section \ref{sec:su2}.
Because of the semisimplicity of $SU(N) $, all parameterizations lead to a Wei-Norman formula that is subject to singularities and as such $ \Xi^{-1} $ has only a local validity.
Call $ \Sigma $ the singular set of $\Xi $.
By inverting $ \Xi $ when possible, equation \eqref{eq:matrix-xi} assumes the more traditional aspect of a system of first order differential equations in the $ \gamma_i $ variables:
\begin{equation}
 \begin{bmatrix} \dot \gamma_1 \\ \vdots \\ \dot \gamma_n \end{bmatrix} = \Xi (\gamma_1 , \ldots , \gamma_n )^{-1}  \begin{bmatrix} a_1 + u_1 \\ \vdots \\ a_n + u_n \end{bmatrix} \qquad \gamma_i(0) = 0 
\label{eq:matrix-xiinv}
\end{equation}
If the time evolution of one of the two vectors of coordinates $ \gamma_i(t) $ or $ a_i + u_i(t) $ is known, the formul{\ae} \eqref{eq:matrix-xi} or \eqref{eq:matrix-xiinv} can be used to obtain the other one.
A method for the explicit closed-form calculation of the Wei-Norman formula is proposed in \cite{Cla-param-diff1} and will not be repeated here.

\subsection{Analysis of the parameter space of the two expansions}

While \eqref{eq:matrix-xi} is global, \eqref{eq:matrix-xiinv} is valid only as long as $ {\rm det}(\Xi)\neq 0 $, and thus the nonsingularity of $\Xi $ needs to be checked at the point of application.  
In general the exponential map does not posses any good property globally: for example for semisimple Lie groups it is surjective but not injective.
The Lie algebra $\mathfrak{su}(N)$ being ``compact'' implies that the parameter space $ \Gamma \subset \mathbb{R}^n $ of the coordinates corresponding to the exponential map, i.e. the set of values of the real coefficients needed to cover the whole Lie group under the exponential map, is a bounded domain, corresponding to the principal values of the logarithm map.
For the single exponential representation and the basis $ A_i $ chosen above, the parameter space is a solid sphere of radius $ 2 \pi $. $ 2 \pi $ is a conventional choice, see \cite{Gilmore1}, p. 127, which originates from the standard choice of a factor $\frac{1}{2} $ in the exponentiation of the Pauli matrices, convention that we also follow in the example of Section \ref{sec:su2}. The radius could obviously be rescaled by any real normalization factor.
In fact, given any vector $ A \in  \mathfrak{su}(N) $, the straight line through the origin $ \gamma A $, $ \gamma \in \mathbb{R} $, is mapped to a one-parameter subgroup (a geodesic curve if the metric is obtained from the Killing form) of $SU(N) $.   
Because of compactness, ${\rm exp}$ is periodic and for such straight line it holds that $ e^{(\gamma + 4 \pi r )A} =   e^{\gamma A} $ for some real constant $ r$ depending on the norm of $ A$.
Since this is true for all $ A = \sum_{j=1}^n \alpha_j A_j $, the principal values of the logarithm map will correspond to a solid ellipsoid in general, and to a solid sphere in $\mathbb{R}^n $ of radius $ 2 \pi $ when the $A_j $ are suitably normalized and the $ \alpha_j $ are such that $\sum_{j=1}^n \alpha_j ^2 =1 $.

The lack of global properties of the exponential map is amplified in the products of exponentials such as \eqref{eq:WeiN-exp1}.
Call $ {\rm pexp} $ the homomorphism
\begin{equation}
\begin{split}
 {\rm pexp} \; : \; \Gamma & \to SU(N) \\
\gamma & \mapsto \prod_{k=1}^n e^{\gamma_k A_{\rho(k)}}
\end{split}
\label{eq:pexp1}
\end{equation}
where $ \rho(k) $ represents the ordering of the basis elements corresponding to the parameterization chosen for $SU(N) $ (in \eqref{eq:WeiN-exp1} $ \rho(k) = k $, in the ZYZ parameterization of $SU(2)$ used below $ \rho(1,\, 2,\, 3) = 3 ,\, 2, \, 3$).
Although we have $ \gamma_k \in \mathbb{R}$, also in this case the parameter space, i.e. the subset of $ \mathbb{R}^n $ containing the principal values of $ {\rm pexp}^{-1} $, is a bounded domain in $\mathbb{R}$.
However, since each $ \gamma_k $ enters into $ {\rm pexp} (\gamma ) $ independently from $ \gamma_j $, $ k\neq j $, this time the parameter space $ \Gamma_{\rm p} \subset \mathbb{R}^n $ corresponds to a cube in $ \mathbb{R}^n $ centered in the origin and with sides of length $ 4 \pi $ for the same basis $ A_j $ for which the logarithm has domain of principal values equal to the sphere of radius $ 2 \pi $.

\section{Elementary quantum gates as output of a system of differential equations}
Calling $ \gamma = [ \gamma_1 \ldots \gamma_n ]^T $, $ a=[a_1 \ldots a_n ]^T $ and $ u =[u_1 \ldots u_n ]^T $, from \eqref{eq:matrix-xiinv} and \eqref{eq:WeiN-exp1} the state of the quantum system can be thought of as the output of the following system of first order nonlinear differential equations: 
\begin{equation}
\begin{cases}
\dot \gamma &= \Xi^{-1} (\gamma) \left( a + u \right) \qquad \gamma(0) =0 \qquad \quad \gamma \in \Gamma_{\rm p} \setminus \Sigma \\
\ket{\psi} & = {\rm pexp} (\gamma)  \ket{\psi_0 } 
\end{cases}
\label{eq:sys-prod-exp}
\end{equation}
While the differential equation \eqref{eq:schrod2} was linear in $U(t) $ and affine in the controls, the differential equation \eqref{eq:sys-prod-exp} is still affine in control but it is highly nonlinear in the variables $\gamma$.

The study of systems of nonlinear differential equations like \eqref{eq:sys-prod-exp} is quite common in systems theory, provided we call $ \gamma $ the ``state vector'' and $ \ket{\psi} $ the ``output''.
As they are quite confusing in the present context, these (and other) systems theory notations will always be recalled between quotes.

\subsection{Coherent states and $\gamma$-coherent states}
In the whole paper we will always neglect global phase factors $ e^{i \phi }$ as they are irrelevant for our purposes: $ \ket{\psi} = e^{i \phi }\ket{\psi} $.

In general, an action of a Lie group on a manifold is said transitive if each pair of elements of the manifold can be joined by an element of the Lie group.
On $ \mathbb{S}^{N-1} $, the Lie group $SU(N) $ acts transitively: $ \forall \; \ket{\psi}, \, \ket{\psi_0} \in \mathbb{S}^{N-1} $ there exists a unitary transformation $ U \in SU(N) $ such that $ \ket{\psi} = U \ket{\psi_0 } $.
The manifold in this case is a homogeneous space of the Lie group.

The following definitions are standard for a quantum mechanical system (see \cite{Zhang1,Perelomov1}).
The {\em maximal isotropy subgroup} of $ \ket{\psi_0 } $ is given by
\[
H_{\ket{\psi_0}} = \left\{ h \in SU(N) \;  \text{ s.t. }  \; h \ket{\psi_0 } = \ket{\psi_0} \right\} 
\]
Since the action is transitive, $ H_{\ket{\psi_0}} $ is a subgroup. In fact, if $ h_1 , \,  h_2 \in  H_{\ket{\psi_0}} $, then $ h_2 h_1 \ket{\psi_0 } = h_2 \ket{\psi_0 } = \ket{\psi_0 } $ and $ h_1^{-1} \ket{\psi_0 } = \ket{\psi_0} $.
States that differ by an element $H_{\ket{\psi_0}} $ are {\em indistinguishable} from $\ket{\psi_0}$: if $U \in H_{\ket{\psi_0}}$ then $\ket{\psi}= U \ket{\psi_0 }= \ket{\psi_0 } $, again (and always) up to an irrelevant global phase factor.
Calling $ C_{\ket{\psi_0}} $ the {\em coset space} of $ \ket{\psi_0} $ in $SU(N)$, i.e. the space of equivalence classes determined by the isotropy subgroup,
\begin{equation}
C_{\ket{\psi_0}} = SU(N) / H_{\ket{\psi_0}}
\label{eq:coset-sp1}
\end{equation}
each $ U \in SU(N)$ can be decomposed into $ U = \Omega \, h $ with $ \Omega \in C_{\ket{\psi_0}} $ and $ h \in  H_{\ket{\psi_0}}$.
The ``effective'' change in $ \ket{\psi_0 } $ given by a unitary transformation $ U =\Omega \, h $ corresponds to the coset component alone: 
\[
\ket{\psi} = U \ket{\psi_0} = \Omega \, h \ket{\psi_0} =  \Omega  \ket{\psi_0}
\]
The set 
\begin{equation}
\left\{ \Omega \ket{\psi_0 } \; \text{ s.t. }  \; \Omega \in C_{\ket{\psi_0}} \right\} = C_{\ket{\psi_0}} \ket{\psi_0 } 
\label{eq:coher-su2}
\end{equation}
is called the set of {\em coherent states} of $\ket{ \psi_0 }$.
As can be seen from \eqref{eq:coher-su2}, the coherent states of $ \ket{\psi_0 } $ are obviously in 1-1 correspondence with the elements of the coset space $  C_{\ket{\psi_0}} $.

For the product of exponentials map \eqref{eq:pexp1}, the preimage $ {\rm pexp} ^{-1} (U) $ may vary from point to point. 
In particular, $  {\rm pexp} ^{-1} (I) $ varies according to the order $ \rho(k) $ selected.
For example, in the canonical coordinates of the second kind ($ \rho(k) = k $) $  {\rm pexp} ^{-1} (I) = \left\{ 0 \right\} $, but in general it might be a nontrivial set.
Call such set $ H_\gamma (U) $, the $\gamma$-isotropy subgroup at $U$ of the map $ {\rm pexp }$ at $U\in SU(N) $
\[
H_\gamma (U) = \left\{ \gamma \in \Gamma_{\rm p} \, \text{ s.t. } \, {\rm pexp} (\gamma ) = U \right\} = {\rm pexp}^{-1} (U) 
\]
For example, $ H_\gamma (I) $ is the set of all coordinates $ \gamma \in \Gamma_{\rm p} $ that certainly do not produce any effect on $ \ket{\psi} $ because their corresponding unitary operator is the identity.
More generally, all $ \gamma $ such that $ {\rm pexp}(\gamma) \in H_{\ket{\psi_0}} $ will produce no effect on an initial quantum state $ \ket{\psi_0 }$.
We indicate the composition of the two types of isotropy subgroups as 
\begin{equation}
H_\gamma (\ket{\psi_0}) =``H_\gamma \circ H_{\ket{\psi_0}} ``= \left\{ \gamma \in \Gamma_{\rm p} \, \text{ s.t. } \, {\rm pexp} (\gamma ) =  h , \, \text{ with }  h \in H_{\ket{\psi_0}} \right\} 
\end{equation}
and call $ H_\gamma (\ket{\psi_0}) $ the {\em $\gamma$-isotropy subgroup at $ \ket{\psi_0 }$}.

In system theory terminology, $ H_\gamma (\ket{\psi_0}) $ should be called the ``unobservable subspace'' of the ``state space'', because ``states'' $ \gamma\in H_\gamma (\ket{\psi_0}) $ are indistinguishable when looked from the ``output'' $ \ket{\psi}$.

Similarly to \eqref{eq:coset-sp1}, one can define the {\em $\gamma$-coset space} $ C_\gamma ( \ket{\psi_0 })$ as
\[
 C_\gamma ( \ket{\psi_0 }) = \Gamma_{\rm p} / H_\gamma (\ket{\psi_0}) 
\]
and call the {\em $\gamma $-coherent states} the values assumed by $ \ket{\psi} $ for $\gamma $ varying in $  C_\gamma ( \ket{\psi_0 })$.

\subsection{Universality of the gates for products of exponentials coordinates}
The universality of the gates property mentioned in the introduction corresponds to $  C_\gamma ( \ket{\psi_0 }) $ being all of $\mathbb{S}^{N-1} $.
It holds almost everywhere and depends on the free Hamiltonian (i.e. on the parameters $ a_i $) being nondegenerate, see \cite{Cla-contr-root1}.
Notice that in the present framework the correct system theoretic concept to use is ``output'' controllability, not ``state'' controllability.
Notice, furthermore, that whenever $ \Sigma \subseteq  H_\gamma (\ket{\psi_0}) $, the singular points of the ``state'' space representation \eqref{eq:sys-prod-exp} are ininfluent in the quantum state manipulation, i.e. they are not seen from the ``output'' $\ket{\psi} $ of \eqref{eq:sys-prod-exp}.
Thus we have the following result:
\begin{proposition}
\label{prop:univ1}
Assume that the system \eqref{eq:schrod1} is controllable and that a fixed order of infinitesimal generators is chosen in \eqref{eq:pexp1}. For a quantum state $\ket{\psi_0}$, 
\begin{itemize}
\item[ i) ] if $ \Sigma \subseteq  H_\gamma (\ket{\psi_0}) $, then the gates in \eqref{eq:pexp1} are universal; 
\item[ ii) ] if $ \Sigma \nsubseteq H_\gamma (\ket{\psi_0}) $, then the gates in \eqref{eq:pexp1} may not be universal. In this case the analysis of the singular locus $ \Sigma $ is required in order to establish universality of the gates.
\end{itemize}
\end{proposition}
While the value of $\Sigma $ depends only on the order chosen in \eqref{eq:pexp1}, $  H_\gamma (\ket{\psi_0}) $ varies also with the initial condition $ \ket{\psi_0 } $.
Thus the previous sufficient condition for universality of the gates may hold only for some values of $ \ket{\psi_0 }$.
Notice that, provided that \eqref{eq:schrod1} is controllable, the sufficient conditions above does not depend on the control authority available (i.e. on how many $ u_j $, $ j=1, \ldots , n $ are different from $0$).
However, when $ \Sigma \nsubseteq H_\gamma (\ket{\psi_0}) $ and the singular points of $\Sigma $ are investigated, universality depends essentially on which control fields are available.

All the constructions and characterizations discussed so far are easily clarified by means of a concrete example.

\section{Application to a single qbit}
\label{sec:su2}
The most elementary quantum state of interest here is the single qbit, defined on the sphere $\mathbb{S}^1 $.
In the computational basis $ \ket{0} $, $ \ket{1}$, the state is $ \ket{\psi}\simeq \begin{bmatrix} y_1 & y_2 \end{bmatrix}^T $ with $ |y_1|^2 + |y_2|^2 = 1 $, thus the complex sphere in $ \mathbb{C}^2 $ is expressed in the computational basis coordinates as
\[
\mathbb{S}^1 = \left\{ \begin{bmatrix} y_1 \\ y_2 \end{bmatrix} \in \mathbb{C}^2  \;  \text{ s.t. }  \; |y_1 |^2 + |y_2|^2 =1 \right\}
\]
$SU(2)$ acts transitively on $ \mathbb{S}^1 $ and we shall concentrate on $ SU(2)$ unitary transformations as elementary gates for $ \ket{\psi}$.
A skew-symmetric basis for $ \mathfrak{su}(2) $ is obtained from the Pauli matrices
\[
\sigma_1 =  \begin{bmatrix} 0 & 1 \\ 1 & 0 \end{bmatrix} \quad
\sigma_2 =  \begin{bmatrix} 0 & -i \\ i & 0 \end{bmatrix} \quad
\sigma_3 =  \begin{bmatrix} 1 & 0 \\ 0 & -1 \end{bmatrix}
\]
for example by taking $ A_j =\frac{i}{2} \sigma_j $, $j=1,\,2,\,3$, i.e.
\begin{equation}
A_1 = \frac{1}{2} \begin{bmatrix} 0 & i \\ i & 0 \end{bmatrix} \quad
A_2 = \frac{1}{2} \begin{bmatrix} 0 & 1 \\ -1 & 0 \end{bmatrix} \quad
A_3 = \frac{1}{2} \begin{bmatrix} i & 0 \\ 0 & -i \end{bmatrix}
\label{eq:basis-su2}
\end{equation}
and it corresponds to all real structure constants $ c_{12}^3 = c_{23}^1 = c_{31}^2 = 1$.
The corresponding adjoint matrices are
\[
{\rm ad}_{A_1} =  \begin{bmatrix}
0 & 0 & 0 \\
0 & 0 & -1 \\
0 & 1 & 0 
\end{bmatrix} \quad
{\rm ad}_{A_2} =  \begin{bmatrix}
0 & 0 & 1 \\
0 & 0 & 0 \\
-1 & 0 & 0 
\end{bmatrix} \quad
{\rm ad}_{A_3} =  \begin{bmatrix}
0 & -1 & 0 \\
1 & 0 & 0 \\
0 & 0 & 0 
\end{bmatrix} \quad
\]
whose exponentials are already known from the literature to be:
\[
e^{\gamma_1 {\rm ad}_{A_1} } =
\begin{bmatrix}
1 & 0 & 0 \cr 
0 & \cos \gamma_{1} & -\sin \gamma_{1} \cr 
0 & \sin \gamma_{1} & \cos \gamma_{1} \cr  
\end{bmatrix} \quad 
e^{\gamma_2 {\rm ad}_{A_2} } =\begin{bmatrix}
\cos \gamma_{2} & 0 & \sin \gamma_{2} \cr 
0 & 1 & 0 \cr 
-\sin \gamma_{2} & 0 & \cos \gamma_{2} \cr  
\end{bmatrix} \quad 
e^{\gamma_3 {\rm ad}_{A_3} } =\begin{bmatrix}
\cos \gamma_{3} & -\sin \gamma_{3} & 0 \cr
 \sin \gamma_{3} & \cos \gamma_{3} & 0 \cr 
0 & 0 & 1 \cr 
\end{bmatrix}
\]
The parameter space $ \Gamma $ i.e. the portion of the Lie algebra that is needed to parameterize the group under the single exponential is a solid sphere of radius $ 2 \pi $ and with all the points on the surface identified (they all correspond to $ -I $, see \cite{Gilmore1}, p. 123).

\subsection{ZYZ quantum logic gates}
Probably the most common set of gates for the single qbit is given by the ZYZ Euler angles. 
Choosing such a parameterization for $SU(2) $, the ZYZ-operations on $ \ket{\psi_0}$ are described by the following product of exponentials
\begin{equation}
\ket{\psi}= U(\gamma_1, \, \gamma_2 , \, \gamma_3 ) \ket{\psi_0 } =  e^{i \gamma_1 \frac{\sigma_3}{2} }  e^{i \gamma_2 \frac{\sigma_2}{2} } e^{i \gamma_3 \frac{\sigma_3}{2} } \ket{\psi_0}=  e^{ \gamma_1 A_3}  e^{ \gamma_2 A_2}  e^{ \gamma_3 A_3}  \ket{\psi_0}
\label{eq:prod-ZYZ1}
\end{equation}
where $ \gamma = \begin{bmatrix} \gamma_1 & \gamma_2 & \gamma_3 \end{bmatrix}^T $ is defined in $\Gamma_{\rm p} = ( - 2 \pi , \, 2 \pi ]^3 $ and $\gamma_i = \gamma_i (t) $.

From the expressions for the exponentials of the Pauli matrices:
\begin{eqnarray*}
e^{\gamma_1 A_1} & = &
 \cos \frac{\gamma_1}{2} I + i \sin \frac{\gamma_1}{2} \sigma_1 = 
\begin{bmatrix}
\cos \frac{\gamma_{1}}{2} &  i\sin \frac{\gamma_{1}}{2}  \cr 
i \sin \frac{\gamma_{1}}{2} & \cos \frac{\gamma_{1}}{2}   
\end{bmatrix} \\
e^{\gamma_2 A_2}  & = &
\cos \frac{\gamma_2}{2} I + i \sin \frac{\gamma_2}{2} \sigma_2 
= \begin{bmatrix}
\cos \frac{\gamma_{2}}{2} & \sin \frac{\gamma_{2}}{2} \cr 
- \sin \frac{\gamma_{2}}{2} &  \cos \frac{\gamma_{2}}{2}   
\end{bmatrix} \\
e^{\gamma_3A_3} & = & 
\cos \frac{\gamma_3}{2} I + i \sin \frac{\gamma_3}{2} \sigma_3 
=\begin{bmatrix}
e^{ i \frac{\gamma_{3}}{2}} & 0 \cr
 0 & e^{-i \frac{ \gamma_{3}}{2}} \cr 
\end{bmatrix}
\end{eqnarray*}
the ZYZ product of exponentials is
\begin{equation}
 U(\gamma_1, \, \gamma_2 , \, \gamma_3 ) = \begin{bmatrix} 
e^{i ( \frac{\gamma_1}{2} + \frac{\gamma_3}{2} ) } \cos \frac{\gamma_2}{2} &  
e^{i ( \frac{\gamma_1}{2} - \frac{\gamma_3}{2} ) } \sin \frac{\gamma_2}{2} \\
-e^{i (- \frac{\gamma_1}{2} + \frac{\gamma_3}{2} ) } \sin \frac{\gamma_2}{2} &
e^{-i ( \frac{\gamma_1}{2} + \frac{\gamma_3}{2} ) } \cos \frac{\gamma_2}{2} 
\end{bmatrix}
\label{eq:U-ZYZ}
\end{equation}

\subsection{Wei-Norman formula for the $ZYZ $ Euler angles}
The Wei-Norman formula corresponding to the ZYZ ordered product \eqref{eq:prod-ZYZ1} is (see \cite{Cla-param-diff1}):
\begin{equation}
\Xi =
 \begin{bmatrix} 0 & - \sin  \gamma_1  & \cos  \gamma_1  \sin  \gamma_2  \\
0 & \cos  \gamma_1 & \sin \gamma_1  \sin  \gamma_2  \\
1 & 0 & \cos \gamma_2 
\end{bmatrix}
\label{eq:wei-n-ZYZ}
\end{equation}
and its inverse
\begin{equation}
\Xi^{-1} = 
 \begin{bmatrix} - \cos  \gamma_1   \cot  \gamma_2 & - \sin  \gamma_1  \cot  \gamma_2  & 1 \\
-  \sin  \gamma_1 & \cos  \gamma_1  & 0 \\
\cos  \gamma_1  \csc \gamma_2 & \sin \gamma_1 \csc \gamma_2  & 0 
\end{bmatrix}
\label{eq:wei-ninv-ZYZ}
\end{equation}
Since 
\[
 \det(\Xi ) = \sin \gamma_2
\]
the singular points correspond to $  \gamma_2 = k \pi $, $ k \in \mathbb{Z} $, as is well-known for such a parameterization.
Thus, in $\Gamma_{\rm p} $, $ \Xi^{-1}$ can be used everywhere except in $\Sigma = \left\{ \gamma \in \Gamma_{\rm p} \; \text{s.t.} \; \gamma_2 =-\pi, \, 0 , \, \pi , \, 2 \pi  \right\} $.

It is worth emphasizing that it is a fundamental topological fact that singularities cannot be avoided in a minimal parameterization of a semisimple Lie group.
One possible way to get around the problem is obviously to use ``redundant'' parameterizations like quaternions, another to choose two different local charts i.e. two sets of of exponential coordinates corresponding for example to a different ordering in \eqref{eq:prod-ZYZ1}.

\subsection{Example: nuclear spin qbit}
In \eqref{eq:sys-prod-exp}, some of the $ a_j $ and/or $ u_j $ might be zero.
If for example the free Hamiltonian $ H_0 $ is diagonal then $ a_1 = a_2 =0 $.
In this case, most likely $ u_3 = 0 $ if the dipole approximation for the coupling with the control fields is considered.
Assume that independent fields along both X and Y directions are available.  
Using \eqref{eq:wei-ninv-ZYZ}, \eqref{eq:sys-prod-exp} reads as
\begin{equation}
\begin{cases}
\begin{bmatrix} \dot \gamma_1 \\ \dot \gamma_2 \\ \dot \gamma_3 \end{bmatrix}
=\begin{bmatrix} a_3 \\ 0 \\ 0 \end{bmatrix} 
+ \begin{bmatrix} - \cos  \gamma_1   \cot  \gamma_2 & - \sin  \gamma_1  \cot  \gamma_2   \\
-  \sin  \gamma_1 & \cos  \gamma_1   \\
\cos  \gamma_1  \csc \gamma_2 & \sin \gamma_1 \csc \gamma_2  
\end{bmatrix}
\begin{bmatrix} u_1 \\ u_2 \end{bmatrix}  
\\ \\
\ket{\psi} \simeq \begin{bmatrix} y_1 \\ y_2 \end{bmatrix}  = 
 \begin{bmatrix} 
e^{i ( \frac{\gamma_1}{2} + \frac{\gamma_3}{2} ) } \cos \frac{\gamma_2}{2} &  
e^{i ( \frac{\gamma_1}{2} - \frac{\gamma_3}{2} ) } \sin \frac{\gamma_2}{2} \\
-e^{i (- \frac{\gamma_1}{2} + \frac{\gamma_3}{2} ) } \sin \frac{\gamma_2}{2} &
e^{-i ( \frac{\gamma_1}{2} + \frac{\gamma_3}{2} ) } \cos \frac{\gamma_2}{2} 
\end{bmatrix}\begin{bmatrix} y_{1_0} \\ y_{2_0} \end{bmatrix}
\end{cases}
\label{eq:mod-spin1/2}
\end{equation}
The model \eqref{eq:mod-spin1/2} corresponds to the well-known example of a single nuclear spin $ \frac{1}{2} $ system immersed in a magnetic field which is constant along the Z axis and varying in the X and Y directions.

\subsection{Coherent states for $SU(2)$}

The isotropy subgroup of a generic state $ \ket{\psi}$ can be obtained by solving the algebraic equation
\begin{equation}
\ket{\psi} = U(\gamma_1, \, \gamma_2 , \, \gamma_3 )\ket{\psi}
\label{eq:isotropy-ZYZ}
\end{equation}
The ZYZ representation is convenient is this respect since it is easy to notice that \eqref{eq:isotropy-ZYZ} is solved for all $ \ket{\psi_0}$ at least by the submanifold $ \gamma_2 = 0 $ and $ \gamma_3 = -\gamma_1$.
In fact $ U( \gamma_1 , \, 0 , \, - \gamma_1 ) = I $ and therefore $ H _\gamma (\ket{\psi_0 })  \supseteq H_\gamma (I) = \left\{ \gamma \in \Gamma_{\rm p} \; \text{s.t.} \; \gamma_2 =0 , \, \gamma_3 = - \gamma _1  \right\}$.
A complete computation of the $ H_\gamma (\ket{\psi_0}) $ depends on the value of $ \ket{\psi_0 } $.
For example, for $ \ket{0} \simeq [ 1 \; 0 ]^T $  \eqref{eq:isotropy-ZYZ} is solved by all $ \gamma $ such that $ \gamma_2 =0 $ or $ \gamma_2 = 2 \pi $, since in this case $ e^{ i ( \frac{\gamma_1}{2} + \frac{\gamma_3}{2} ) }$ becomes an ininfluent global phase factor: $ \ket{0} =\pm e^{ i ( \frac{\gamma_1}{2} + \frac{\gamma_3}{2} ) } \ket{0} $.
Thus $  H_\gamma   (\ket{0}) =  \left\{ \gamma \in \Gamma_{\rm p} \, \text{ s.t. } \gamma_2 = 0, \, 2 \pi  \right\} $.
For $\gamma_2 =0 $ and $\frac{\gamma_1} {2} + \frac{\gamma_3}{2} = \frac{\pi}{4} $, this corresponds to the so-called ``Z gate'': $ U(\gamma_1 , \, 0 , \frac{\pi}{2} - \gamma_1 ) = e^{\gamma_1 A_3 }e^{(\frac{\pi}{2} - \gamma_1) A_3 }=  \begin{bmatrix} 1 & 0 \\ 0 & -1 \end{bmatrix} $.
In fact $ \ket{0} $ is a highest weight state in $ \mathbb{S}^1 $, for which the computation of coherent states is easier \cite{Zhang1}.
If instead  $ \ket{\psi_0 }$ is a superposition ($ y_{1_0} \neq 0 $ and $  y_{2_0} \neq 0 $), then the $\gamma$-isotropy subgroup contains $H_\gamma (I) $ but may also contain other points, according to the value of $ \ket{\psi_0 } $.
We are in the case ii) of Proposition \ref{prop:univ1}: $ \Sigma \nsubseteq H_\gamma (\ket{\psi_0}) $ for some $ \ket{\psi_0 } $ (other than $ \ket{0} $ and $ \ket{1} $).

\subsection{Analysis of the singularity of the Wei-Norman formula}
In $\Sigma $, the Wei-Norman formula \eqref{eq:wei-ninv-ZYZ} is not valid anymore. 
However, in this case \eqref{eq:wei-n-ZYZ} becomes:
\begin{equation}
\Xi (\gamma_1 , \, 0 , \, \gamma_3 ) =
 \begin{bmatrix} 0 & - \sin  \gamma_1  & 0 \\
0 & \cos  \gamma_1 & 0 \\
1 & 0 & 1 
\end{bmatrix}
\label{eq:wei-n2-ZYZ}
\end{equation}
Since in $ \Sigma $, ${\rm rank} \, \Xi(\gamma_1 , \, \gamma_2  , \, \gamma_3 ) = 2 $, varying $u$ at most two of the $ \dot{\gamma }_i $ can be varied independently.
From \eqref{eq:wei-n2-ZYZ} and \eqref{eq:sys-prod-exp} we get 
\begin{eqnarray}
\dot{\gamma}_1 + \dot{\gamma}_3 & = & a_3 + u_3 
\label{eq:gamma13dyn-1}  \\
\dot{\gamma}_2 & = & -(a_1 + u_1 ) \sin \gamma_1 + ( a_2 + u_2 ) \cos \gamma_1 
\label{eq:gamma2dyn-1}
\end{eqnarray}
i.e. it is still possible to obtain a well-defined expression for the dynamics at which $\gamma_2$ obeys.

In particular, in the example of \eqref{eq:mod-spin1/2}, the system in $\Sigma $ reduces to 
\[
\begin{cases}
\begin{bmatrix} \dot{\gamma}_1 + \dot{\gamma}_3 \\ \dot \gamma_2 \end{bmatrix}
& = \begin{bmatrix} a_3  \\ - u_1  \sin \gamma_1 +  u_2  \cos \gamma_1  \end{bmatrix}   \\ \\
\qquad  \begin{bmatrix} y_1 \\ y_2 \end{bmatrix}  & = 
 \begin{bmatrix} 
e^{i ( \frac{\gamma_1 + \gamma_3}{2} ) } y_{1_0}  \\
e^{-i ( \frac{\gamma_1 + \gamma_3}{2} ) } 
 y_{2_0} \end{bmatrix}
\end{cases}
\]
from which it is easy to observe that the free Hamiltonian only induces a relative phase transformation on a superposition state $\ket{\psi_0 }$ (as expected, $H_0 $ being diagonal), while the actuated system can always be steered out of the ``unobservable subspace'' $ H_\gamma (\ket{\psi_0}) $ (and also of $ \Sigma $) by means of the available control authority $u_1 $ and $u_2 $.

In summary: almost all the singular points of $ \Xi $ are ``unobservable'' for $ \ket{\psi} $, i.e. they produce no influence on the quantum state.
Some care is required in the algebraic set $ \Sigma $ because the exponential coordinates are not defined.
The problem is analogous to the lack of global coordinates on a manifold.
Even if $ \Xi^{-1} $ is not defined in $\Sigma $, the implicit equation \eqref{eq:matrix-xi} still governs the dynamics and can be used to pull $\gamma $ out of $\Sigma $.

\section{Conclusion and outlook}
The aim of this work is to build the differential equations of a driven quantum dynamics in the formalism best suited for quantum computation, where a predefined set of quantum gates can be expressed directly in terms of the physical control parameters via exponential coordinates on the group of transformation.
Beside the general remark that choosing coordinates on a manifold is always necessary when one wants to do computations, in the present context it is worth stressing that coordinates (here exponential coordinates) have the appealing interpretation of ``gains'' in the logic gates of the quantum hardware.

We would like to emphasize that the method presented here is by no means the only way to relate the continuous and discrete dynamics.
Indeed each implementation proposal for quantum computers has its own method to generate gates, relying on the already existing techniques for state manipulation.
See \cite{Nielsen1} for a survey of the different methods.
The practice on NMR spectroscopy for example, bypasses the synthesis of the control fields in continuous time by applying strong pulses of predetermined time width and by checking the overall effect on the Hamiltonian via coherent averaging and syncronous stroboscopic measurements \cite{Ernst1}.
The use of piecewise constant controls mentioned in the Introduction is indeed another possible solution.
On the contrary, notice that the well-established formalism of sampled (bilinear) control systems, which one could think of using in the present context, is actually not suited, as the potential power of quantum computing resides in the continuous superposition of quantum states which is destroyed by sampling.

The method we are proposing here not only enable to analytically describe the unitary transformations in terms of the available control inputs, but it also gives the possibility of monitoring their variations due to dynamical perturbations or uncertainties in the external fields.
Since we choose to work with exponential coordinates for the unitary group $SU(N)$ of the $N$qbit, non unitary disturbances like decoherence phenomena cannot be represented.
We plan to extend the method to open quantum systems in the future.

 \bibliographystyle{abbrv}


\end{document}